\documentstyle[12pt]{article}

\begin{document}

\begin{flushright}
Preprint CAMTP/94-11\\ December 1994\\
\end{flushright}
\begin{center}
\large {\bf SUPPLEMENT TO THE PAPER:}\\ {\bf  Separating the
regular and irregular energy levels and  their statistics  in Hamiltonian
system with mixed classical dynamics\footnote{Preprint CAMTP/94-10 submitted to
J. Phys. A: Math. Gen.}}\\
\vspace{0.3in}
\normalsize Baowen
Li\footnote{e-mail Baowen.Li@UNI-MB.SI} and Marko Robnik\footnote{e-mail
Robnik@UNI-MB.SI}\\
\vspace{0.2in} Center for Applied Mathematics and
Theoretical Physics,\\ University of Maribor, Krekova 2, SLO-62000 Maribor,
Slovenia\\
\end{center}
\vspace{0.2in}
{\bf Abstract.}
As a technical
supplement to the above mentioned paper we present 192 consecutive eigenstates
for the Robnik billiard with  the shape parameter $\lambda=0.15$ from 10,001st
to 10,192nd, by showing the plots  in the  configuration space and in the phase
space. The latter is smoothed projection of the Wigner function onto the
surface of section. By comparison with the classical SOS plots we thus examine
all eigenstates and classify them  in regular and irregular: There are 70
regular states and 122 irregular states, thus giving the estimate of the
relative measure of the regular component  $\rho_1=0.365$, which is in
excellent agreement with the classical value $\rho_1=0.360$
calculated and reported by Prosen and Robnik (1993).
\\\\

\\\\
Submitted for publication.

\normalsize
\vspace{0.3in}
\newpage

\section*{Explanation of plots}

In this work we provide additional material to our recent paper (Li and Robnik
1994a). The method and the mathematics of analyzing the eigenstates in the
Robnik billiard defined as the quadratic complex conformal map of the unit disk
(in the z-plane, $|z|\le1$) onto the complex w-plane, namely
\begin{equation}
{\cal B}_{\lambda} = \{w|w=z+\lambda z^2,\quad |z|\le 1\}.
\label{eq:RB}
\end{equation}
with $\lambda=0.15$, is presented in our recent paper (Li and Robnik 1994b).
\\\\
The wavefunctions we are looking at are the eigenfunctions of the Schr\"odinger
equation (Helmholtz equation):
\begin{equation}
\Delta \Psi + E \Psi = 0, \qquad \Psi = 0 \quad {\rm at\quad the\quad
boundary\quad of\quad {\cal B}_{\lambda}}
\label{eq: Helm}
\end{equation}
where $E = k^{2}$ is the eigenenergy and $k$ the wavenumber. (So we are using
units such that Planck's constant $\hbar=1$ and $2m=1$, where $m$ is the mass
of the point billiard particle.)
\\\\
In the configurational plots in figures 1-8 we show the contours of constant
probability density for the eigenstates 10,001st through 10,192nd with each
figure containing 24 consecutive plots ordered as left-right top-down
sequence. The scale and the coordinates are uniquely specified by the equation
defining the boundary namely, $w=z+\lambda z^2$, where $|z|=1$. The contours of
constant probability density are chosen in steps of $1/8$ of the maximal value
of each individual plot.
\\\\
In order to investigate the eigenstates in the quantum (Wigner) phase space we
have first to define the classical phase space and the surface of section. The
usual bounce map (Poincar\'e map) in the Birkhoff coordinates (arclength versus
tangent unit velocity vector component) is not suitable for our purpose,
because $\Psi$ vanishes on the boundary. Therefore we choose the surface of
section defined by $v= {\rm Im}(w)=0$:
Our surface of section is now specified by the
crossing point coordinate $u$ on the abscissa versus the conjugate momentum
equal to the tangential component of the velocity vector of length $k$  with
respect to the line of section $v=0$.
In figure 9 we show 24 identical plots of the geometry of
the largest chaotic component for $\lambda=0.15$, in exactly the same size as
the quantal phase space plots which we will show in figures 10-17, which
enables the reader to perform the comparison of classical and quantal plots and
the classification of the eigenstates in regular and irregular, by overlaying
the sheets.
We do not show further details of the KAM scenario inside the stability islands
in order not to obscure the structure of the phase space.
\\\\
The Wigner function (of an eigenstate $\Psi(u,v)$) defined in the full phase
space $(u,v,p_u,p_v)$ is
\begin{equation}
W({\bf q,p}) = \frac{1}{(2\pi)^{2}} \int d^{2}{\bf X}
\exp(-i{\bf p}\cdot{\bf X}) \Psi({\bf q - X}/2)\Psi({\bf q + X}/2)
\label{eq:Wigner}
\end{equation}
where we have specialized to our real $\Psi$ case, and also two degrees
of freedom and $\hbar=1$. Here ${\bf q} = (u,v)$ and ${\bf p} =(p_u,p_v)$.
In order to compare the quantum Wigner functions with the classical Poincar\'e
maps on the surface of section we define the following projection of
(\ref{eq:Wigner}) given as
\begin{equation}
\rho_{SOS}(u,p_u) = \int dp_v W(u,0,p_u,p_v),
\label{eq:proj}
\end{equation}
which nicely reduces the number of integrations by one and is equal to
\begin{equation}
\rho_{SOS} (u,p_u) = \frac{1}{2\pi} \int dx \exp(ixp_u) \Psi(u+\frac{x}{2},
0)\Psi(u-\frac{x}{2},0)
\label{eq:qsos}
\end{equation}
As is well known the Wigner function and its projections are not positive
definite and indeed one typically finds small and inconvenient but
nevertheless physical oscillations around zero which seriously obscure the main
structural features.  Therefore in order to
compare the classical and quantal phase space structure it is advisable to
smooth the Wigner function or its projections (\ref{eq:qsos}) by a normalized
Gaussian kernel with a suitably adapted dispersion. Such procedure has been
introduced and used in (Takahashi 1989, Leboeuf and Saraceno 1990,
Heller 1991, Prosen and Robnik 1993), which is Husimi type representation
but the effective area of our Gaussian kernel will be smaller than  $2\pi$.
\\\\
In figures 10-17 we show the smoothed object (\ref{eq:qsos}) for each
consecutive eigenstate from the 10,001st to the 10,192nd, each figure
containing 24
states arranged in order left-right top-down. The lowest contour is at the
level of $0.15$ of the maximal value and the step size  upwards is also $0.15$
of the maximum.
\\\\
The examination of the states in figures 10-17 and classification leads to the
following result: There are 70 regular and 122 irregular states. This yields
the relative fraction of regular levels $\rho_1=0.365$, in excellent agreement
with the classical value $\rho_1=0.360$ (Prosen and Robnik 1993).

\section*{Acknowledgements}

We thank Toma\v z Prosen for a few computer programms. This research was
supported by the Ministry of Science and Technology of the Republic of
Slovenia.

\section*{References}

Heller E J 1991  in {\em Chaos and Quantum Systems (Proc. NATO ASI Les Houches
Summer School)} eds M-J Giannoni, A Voros and J Zinn-Justin,
(Amsterdam: Elsevier) p547\\\\
Leboeuf P and Saraceno M 1990 {\em J. Phys. A: Math. Gen.} {\bf 23} 1745\\\\
Li Baowen and Robnik M 1994a {\em Preprint CAMTP/94-10},
submitted to {\em J. Phys. A: Math. Gen.} in December\\\\
Li Baowen and Robnik M 1994b {\em Preprint CAMTP/94-8}
submitted to {\em J. Phys. A: Math. Gen.} in October\\\\
Prosen T and Robnik M 1993 {\em J. Phys. A: Math. Gen.} {\bf 26} 5365\\\\
Takahashi K 1989 {\em Prog. Theor. Phys. Suppl. (Kyoto)} {\bf 98} 109\\\\
\end{document}